# Position based dynamic of a particle system: a configurable algorithm to describe complex behaviour of continuum material starting from swarm robotics

**Ramiro dell'Erba**
*ENEA Technical Unit technologies for energy and industry – Robotics Laboratory*

**Abstract**

In a previous work we considered a two-dimensional lattice of particles and calculated its time evolution by using an interaction law based on the spatial position of the particles themselves. The model reproduced the behaviour of deformable bodies both according to the standard Cauchy model and second gradient theory; this success led us to use this method in more complex cases. This work is intended as the natural evolution of the previous one in which we shall consider both energy aspects, coherence with the principle of Saint Venant and we start to manage a more general tool that can be adapted to different physical phenomena, supporting complex effects like lateral contraction, anisotropy or elastoplasticity.

## 1. Introduction

It is well-known that time evolution of a material particles system is determined by Newton's dynamics laws; however in recent years, especially with the evolution of Computer Graphics driven by videogames applications, there has been great interest in studying the evolution of a particle system whose motion is simply determined by their relative position in a frame, without solving the differential equations of dynamics. The name of this method is position based dynamics (PBD) [1], [2]. Such methods, therefore, do not determine forces and solve differential equations but use a position-based approach. The physically-based simulation of deformable solids has been an active research topic of computer graphics for many years: the aim is to simulate the behaviour of real materials to achieve graphically realistic results. In the beginning simulations for videogames applications widely used Continuum Mechanical methods, solving equations by finite element methods (FEM). To obtain robust simulations very small time steps are required by these methods; therefore they cannot be used in interactive situations owing to the large machine time used.
In spite of this, still now, the first approach to simulate deformable objects by continuum mechanics is to discretize equations and to solve them using numerical integration. This can be done in several ways, but many of them are affected by the stability problem, arising from stiff differential equations and can be managed using very small time steps, resulting in high computational costs. In the meantime Graphic Processing Units (GPU) based solvers and adaptive meshes were growing and used to simulate complex behaviour in real-time so PBD methods became popular because they are fast, robust and easily configurable. In PBD knowledge of traditional forces are avoided in favour of position displacements; the problem is, therefore, transformed in a geometric constraint between configurations. The final positions of the particles are determined by minimizing the distance between the reference shape and the deformed shape of a body. This minimization process requires computation of translational vectors for both shapes and of a rotation matrix. Many solutions [3],[4] have been proposed to enhance the efficiency of the method that, owing to the matrix nature, can be parallelized between the cores of GPU to reduce calculation time. The PBD methods result in a physically plausible behaviour of the continuum but suffer from limitations when modelling complex material properties and describing interactions between heterogeneous bodies [5]. In our approach we try to combine the advantages of both the continuum mechanical and the position-based approaches to describe complex physical phenomena, trying to keep the simulation easy to implement and customizable.

In previous works [6-7-8] we have considered a discrete particle system and have determined its evolution by a rules system that calculated the displacement of the particles, in each time step, as functions of their neighbours relative position. Since finite element method (FEM) is a reliable and well-known numerical approach for both classical and generalized continua [see 9-16 for applications], we compared the results with the corresponding classic mechanical continuum case, whose equations had been solved by FEM simulations with good agreement. In this work we intend to generalize the concept to cover some aspects that have been left not discussed previously. The aim of the proposed model was to develop a suitable more general numerical tool capable of modelling the behaviour of deformable bodies, and to take into account higher gradient constituent relations. This approach seemed particularly promising considering the emerging role of micro structured continua, manufactured with computer-aided methods, as a technological resource [see 17-29], because the presence of a complex microstructure often leads to macroscopic behaviours that require generalized continua for their accurate modelling (see [30-39] for more details and [40] for a historical survey on the subject). Moreover, an algorithm based on the geometric centroid of the neighbours of a given particle is consistent with the idea of (locally) minimizing an elastic potential, as the centroid has the well-known properties of minimizing the sum of the squared distances from a set of given points in an Euclidean space. Therefore, the proposed algorithm seems a natural discrete approach from the variational point of view [41-46]. The advantage of using position-based dynamics is in computational simplicity and time machine leading to results similar to those obtainable with FEM but in a much shorter time; it also provides a useful point of view that could be able to help in understanding what features are important in the deformation without solving differential equations. The general purpose of the complete work, at a future point, is to present a fast and robust method to calculate deformations of a three dimensional body of any shape when a part of it is subject to a time dependent displacement that supports complex physical phenomena. The approach tries to combine continuum mechanical material models with a position-based method using an explicit time integration scheme to manage complex physical effects like isotropic and anisotropic elastic behaviour as well as the effects of lateral contraction. This will be achieved without solving dynamic equations but only using a PBD method. To repeat behaviours, described by constitutive equations of the materials (e.g. Poisson effect), we introduce geometric constraint on the lattice and rules ad hoc on the displacements of the points.

We have written a complete customizable and modular algorithm easily expandable to every new feature we would like to introduce. One of the main advantages of the proposed algorithm is the fact that it automatically takes into account large deformation elasticity, which is a topic having an increasing role in today's research [47-53]. In this first paper we are considering two-dimensional problems and two of the five Bravais lattices. In the next works, the method will be extended in order to simulate fluids and more complex structures.

## 2. The origin of the problem

In [54], [55] the author was investigating the calculation of the geometric configuration of submarine swarm robots by the single elements; this is very important because the swarm, like school fish, adapt its configuration depending on the mission assigned. The concept of robot swarms has been a study theme, for the scientific community, for several years. Swarm research has been inspired by biological behaviours, like those of bees [56], [57], [58] for a long time to take advantage by social activities concepts [59] labour division, task cooperation and information sharing. A single-robot approach is affected by failures that may prevent the success of the whole task. On the contrary, a multi-robot approach can benefit from the parallelism of the operation and by the redundancy given by the usage of multiple agents. Moreover the operator has the possibility to have multiple views simultaneously. In a swarm the members operate with a common objective, sharing the job workload; the lack of one member can be easily managed by redistributing the job among the others. This feature is especially useful if we consider application as discovery and surveillance of a submarine area. A swarm can be considered as a single body, offering the advantage of a simple way of interfacing with the human end-users and overcoming the problem of the control of a large number of individuals. In the swarm there is no central brain, mainly because of the excess needs in band pass requested by such a brain. Instead each individual must possess an intelligent local control system capable of managing its choices according to the choices of the neighbours on the basis of the available data. Data coherence am long the swarm, being affected by the position of the member and by the data propagation speed is also a research topic. What makes swarms interesting is their capability to fill and control large volume of water by means of a network of cooperating sensors and their capability to move in the most interesting zones, increasing density where a major need is required. The members geometrical distribution is flexible and adaptable to the task and environmental characteristics. As an example if priority is to maximize exploration volume the swarm has to maintain a great spatial dispersion and communication band pass could be slowed down; conversely, if the priority is around the risk that the amount of exchanged information becomes inadequate to ensure the correct behaviour of the multi body, the system itself can physically react by changing geometry despite of the drop in performances for the assigned task. For this reason, it is of primary importance that a single element of the swarm knows, at least locally, its configuration and can move to reach the desired one. Like birds in nature, the element of the swarm can decide its movements according to what its neighbours are doing. To this end, a positioning and control algorithm has been developed so that it reaches the desired configuration. It was then noted that a quite similar algorithm could adapt to describing PBD problems, because the movement was quite similar to deformation of a viscoelastic body. Therefore, introducing constriction generated by constitutive equations into the relationships describing relative positions between the members of the swarm, we are trying to describe the deformation of a Continuum medium. A numerical tool like this can be useful to describe complex micro-structures, originated from new techniques developed, whose behaviour cannot be investigated by Cauchy Continuum theory and that are generating large quantities of experimental data. The model we are proposing can exhibit a rich range of behaviours just by changing lattice type and these relationships.

## 3. The algorithm

The algorithm, realized by Mathematica (Wolfram Research), to calculate deformation is based on the following steps. The two dimensional continuum body is dicretized in a finite number of particles occupying, in their initial configuration, the nodes of a lattice. The kind of lattice is chosen between the five plane Bravais lattices (see Figure 1); changing lattice we can obtain different results, all other conditions being equal. The object is discretized, see Figure 2 as example. Four kinds of particles are considered, but the modular algorithm is opened to introduce a new kind if required to describe other behaviour; moreover the membership category of the particles can be changed with time during body deformation. The first kind is the leaders, whose motions are assigned, i.e. the imposed strain of the body. Their motion is known and determines the motion of the other particles. The second kinds is the followers, whose motion is calculated by rules involving the motion of other particles and the characteristic of the lattice. This results in a constrained geometrical problem leading to a transformation operator between the matrices representing the particles configuration, $C_t$, for a discrete set of time steps $t_1$, $t_2$, ...$t_n$.... Changing kind of lattice and interaction rules we are able to simulate the behaviour of different constitutive equations materials and reproduce different physical phenomena. By example if to determine the displacement of a particle we consider only its first neighbours (in a chosen metric, as Chebyshev for example) we obtain a first gradient behaviour, while considering also the second shell of neighbours we obtain second gradient effects, see Figure 3. The third kind of particles belongs to the frame. To avoid edge effects, like corners collapse, we surround the body by an external frame of point; a shell, so that any followers look to have the same boundary condition of all the others. This is because the follower position is calculated by a function of its neighbours; therefore it is important that any follower works in the same conditions. Without the frame a corner point has less neighbours, with respect to an internal point; so far if its coordinates are determined, as example, by barycenter of its neighbours this point will be attracted toward the inside and the lattice and will collapse on the other points. The motion of the frame is much simple: It only follows the motion of an assigned follower of its competence; in case the assigned followers are more than one(i.e. in a corner) then an average displacement (or a more generic rule) is computed. The frame can be something more complex than a single shell; as an example if we are considering second gradient interaction we need a double shell to reach

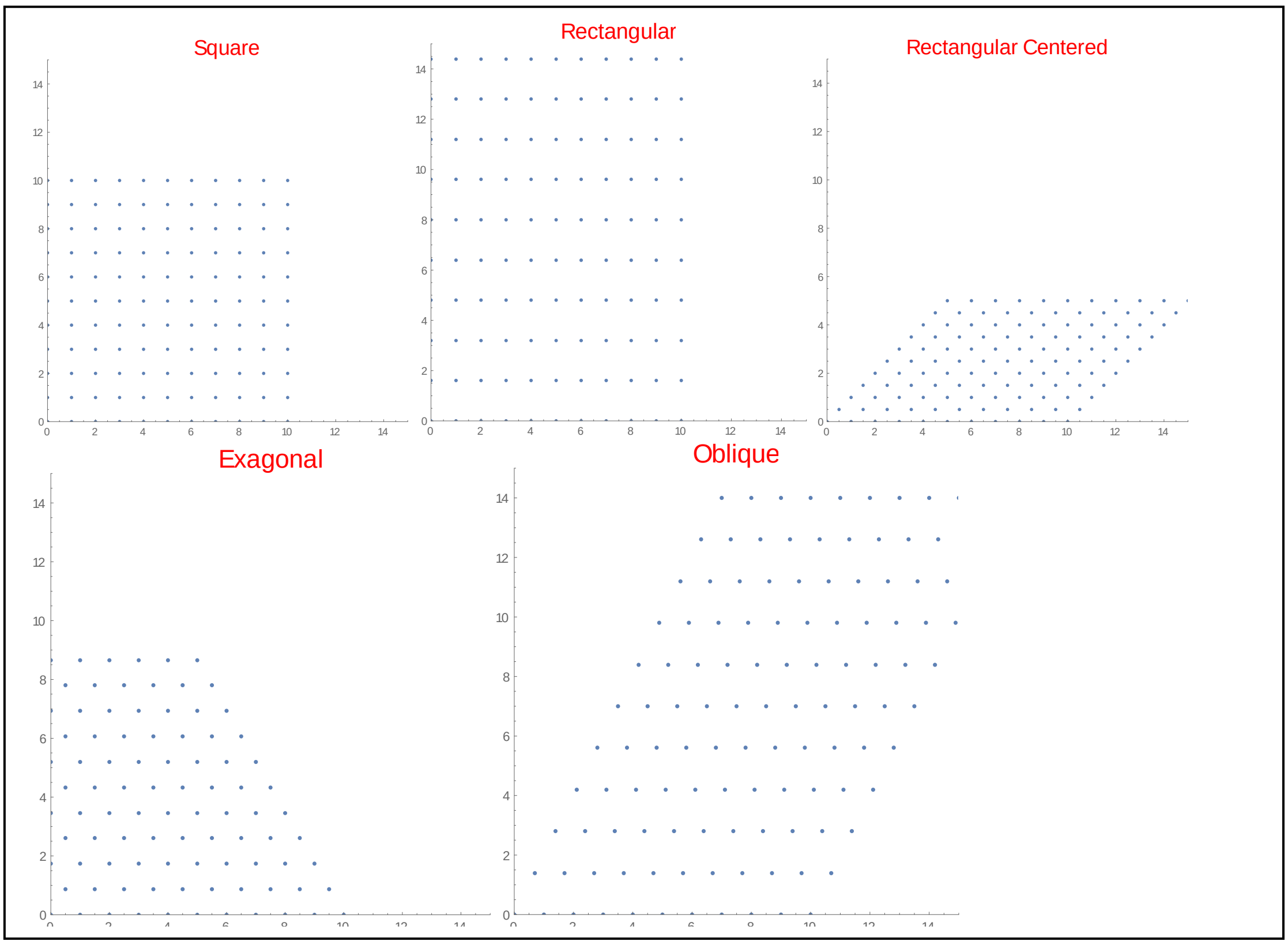

**Figure 1** Bravais plane lattice

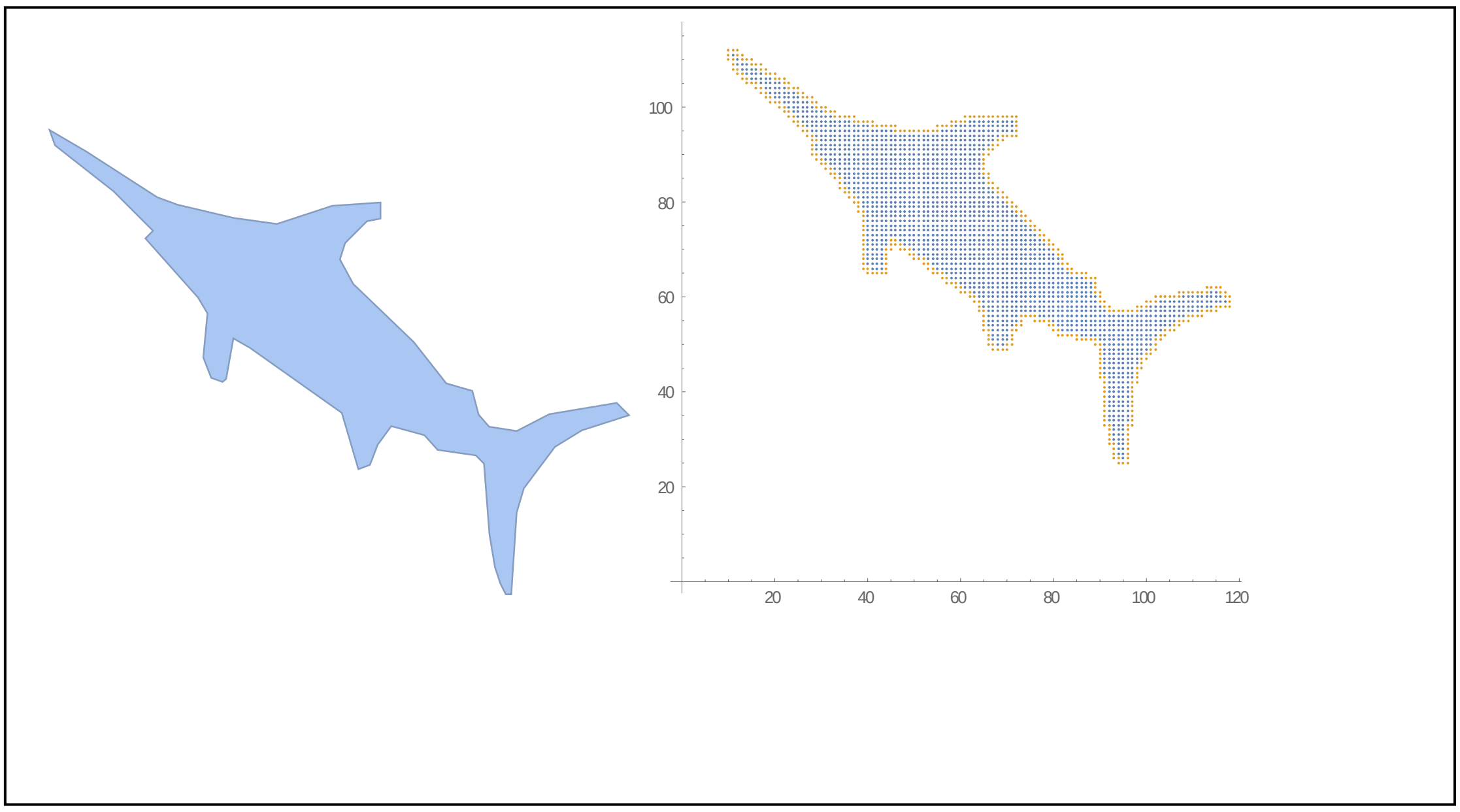

**Figure 2** Example of discretized object in the chosen lattice with frame (yellow points)

our aim that is the homogeneity of the boundary conditions for all the followers. In Figure 3 an example of the three kinds of points is shown, together their first neighbours in the Chebyshev distance ρ, defined in $R^2$ as: $\rho((x_1,x_2,), (y_1,y_2,))=Max(|x_1-y_1|, |x_2-y_2|)$

The last kind is the fictious. They are ghost-like points introduced in some particular case, as to manage fracture; we shall discuss them later.

The processis is the following (seethe flow chart in Figure 4) We choose a two dimensional body. Choose one of the Bravais lattice and discretize the body to obtain a discrete matrix to represent it. We now decide the constrain of the lattice and the interaction rules between the followers, in order to describe the correct behaviour of the constitutive equations of the materials. As an example we can decide that the lattice has no constraints and displacement of a follower point is the average value of the displacements of its first neighbours (first gradient). We build an adequate frame to avoid board effects. We decide the motions of some points, called leaders, for all the time windows we are investigating; we can also decide that they will be leader only for a certain time and late become followers (Category change).

Now we can calculate, for each time step, the new configuration of the lattice in three separate operations. When time increases from $t_0$ to $t_1$ the leaders change their position from initial configuration according to the prescribed equation. So far we build a new intermediate lattice where only the leaders have been moved. Now we take care that the followers are no longer in equilibrium position owing to the leaders displacement. How we can calculate it? As an example if the interactions rule establishes that a follower has to be in the barycenter of all its neighbours we calculate the new position of each follower, taking into account the leader displacement. So far note as at this stage only the leader's neighbours are involved. Finally we take into account the rules governing the frame displacement. This is our lattice at time $t_1$. It is important to note as reached the configuration is not an equilibrium one, because the three operations must be repeated for many time step, after the leaders stop. To be more clear if at time step one the leaders have moved we calculate the followers displacement. This operation involves only the neighbours of the leaders and not the other far followers. Later we calculate the frame displacement to close the loop. Now there are some followers (the neighbours of the leaders neighbours) that there are no longer in equilibrium because there has been the displacement of the leader's neighbours. So we need another time step to adjust the configuration and so on. At a certain time all the followers are involved in the calculation. The followers will suffer the leaders motion after (k-1) time steps where k is the distance from the leaders, measured in layers. In this meaning the leader motion "propagates" through the lattice to influence the position of all the followers in a time depending on the lattice dimensions and how many shell of points are being considered in the neighbours definition. In the same way when leaders stop the followers continue to adjust their position in many time steps. We have often used the rule of centre of gravity to determine followers position that mean:

$$x_j(t) = \frac{\sum_{k=1}^{all\ neighbours\ of\ j} x_k(t)}{N}$$

Where N is the number of neighbours. The same equation is used for the y coordinate.

But we can use different rules in order to approximate different constitutive equations, i.e. we can introduce relative distance between the points into the rule to weight their influence on the followers movement and simulate Hook law, where force is increased with increasing deformation:

$$x_j(t) = \frac{\sum_{k=1}^{all\ neighbours\ of\ j} dis(k,j)x_k(t)}{\sum_{k=1}^{all\ neighbours\ of\ j} dis(k,j)}$$

Where dis(k,j) is the Euclidean distance between the points k and j.

Or we can mix x-y coordinates into the rules so as to make that movement in x direction has effects on the y coordinate (lateral contraction).

$$y_j(t) = K * (x_j(t) - x_j(t_0)) * da + \frac{\sum_{k=1}^{all\ neighbours\ of\ j} y_k(t)}{N}$$

Where da is a function of the distance from the central axes, K a parameter determining the response force and x(t0) the initial x coordinate. This rule leads to the Poisson effect, because we have explained it as an expansion of the x coordinate having influence on the y coordinate.

Moreover we can force the followers movement to overcome the baricenter equilibrium position leading the lattice to oscillate. This will be done in a future paper.

Practically we have a transformation operator between matrices representing initial and final configuration. Take into account that the algorithm is written in a way that, if you like, the neighbours can dynamically change to every time step. The choice to fix the neighbours of every particle at the initial time $t_0$, and to not change them during time evolution of the configurations lies in the desire to imitate a crystalline lattice and therefore to deal with solid phase materials. This means the concept of neighbours is Lagrangian, and neighbourhood is preserved during the time evolution of the system; the only exceptions arise with the fracture algorithm as we can see later. Also the definition of neighbours is customizable by changing metric; as an example we can consider points whose Euclidean distance (weighted or not is another possibility to take into account anisotropies) is less than a threshold or, more physically, the coordination number of the lattice chosen. In the case of second gradient we enlarge the set of points with a supplementary shell.

We also like to introduce a pseudoenergetic consideration to give a contour plot of the strain distribution; in the elastic case we can consider the square distance between the actual configuration $C$ and the reference configuration $C_0$. It must be underlined that this artifice has no direct connection with the usual energy definition (this is the reason we use the term pseudoenergy) but could be useful in understanding deformation. Therefore we introduce two formulations *PE1* and *PE2* for this concept. The first is given by the value, for each step time and in each point describing the configuration, of the sum, extended to the neighbours, of square of the difference between the distances of the point from its neighbours less the distance in the initial configuration i.e.

$$PE1(t,j) = \sum_{k=1}^{all\ neighbours\ of\ j} (dis(t,k,j) - dis(t_0 k,j))^2$$

Where *dis(t,k,j)* is the Euclidean distance between points *k* and *j* at time *t*. This is the formula for the point *j* at time *t*

The reason of this choice lies in the attempt to simulate potential energy of material points subject to Hook law. Be careful with this concept in fracture case; it consider the positions of the real points, not the fictitious, so it still need to be work around.

To compare time contigue configuration $C_t$ and $C_{t-1}$ we define for each point *j* and each time *t*

$$PE2(t,j) = ||C_t - C_{t-1}||$$

Where || is the Norm of the vector defined by the point *j* at time *t* and *t-1*.

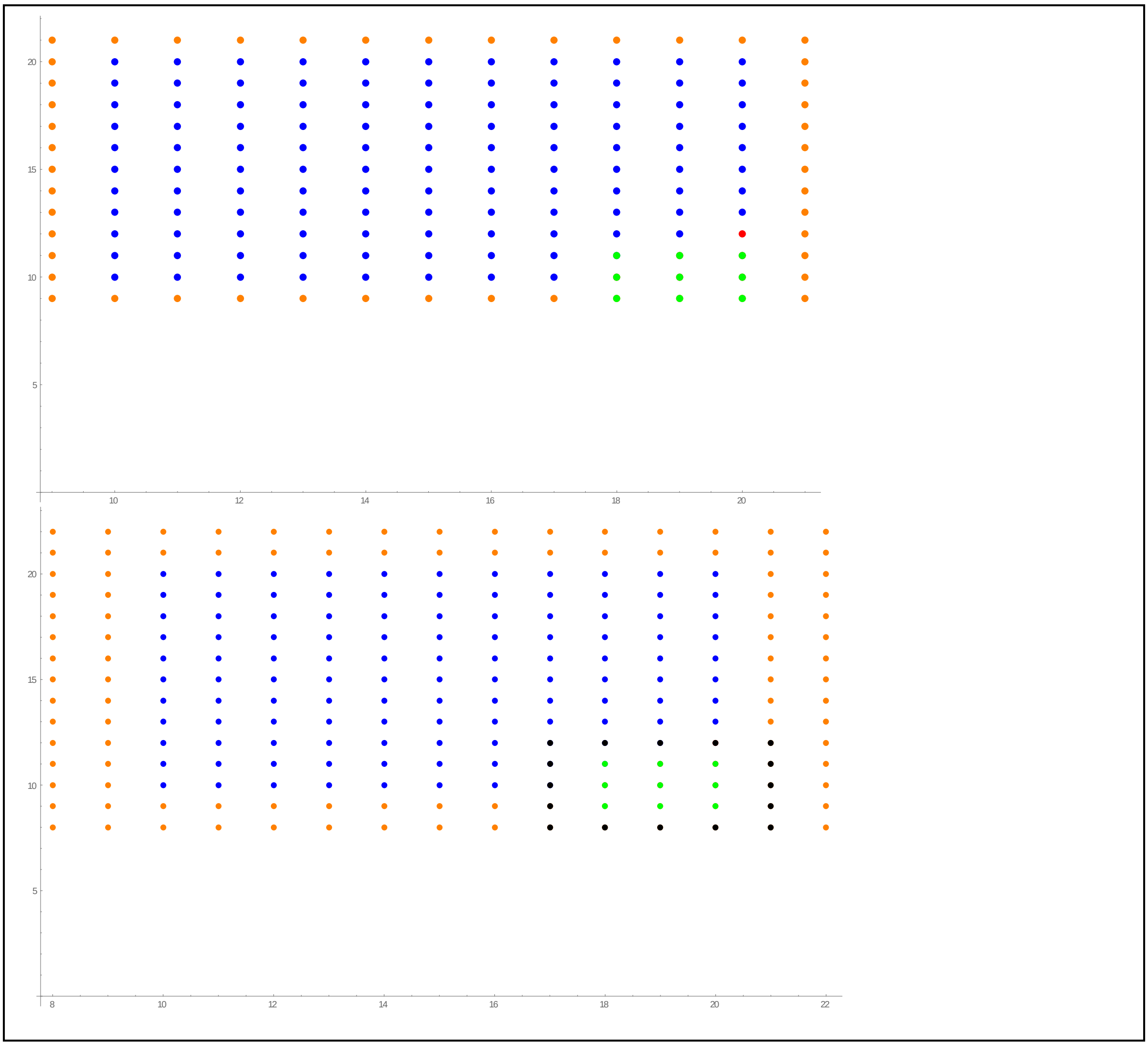

**Figure 3** Example of the three kinds of points. Leaders (red), followers (blue) frame (yellow). An example of first and second neighbours in Chebyshev distance is shown (green points and black)

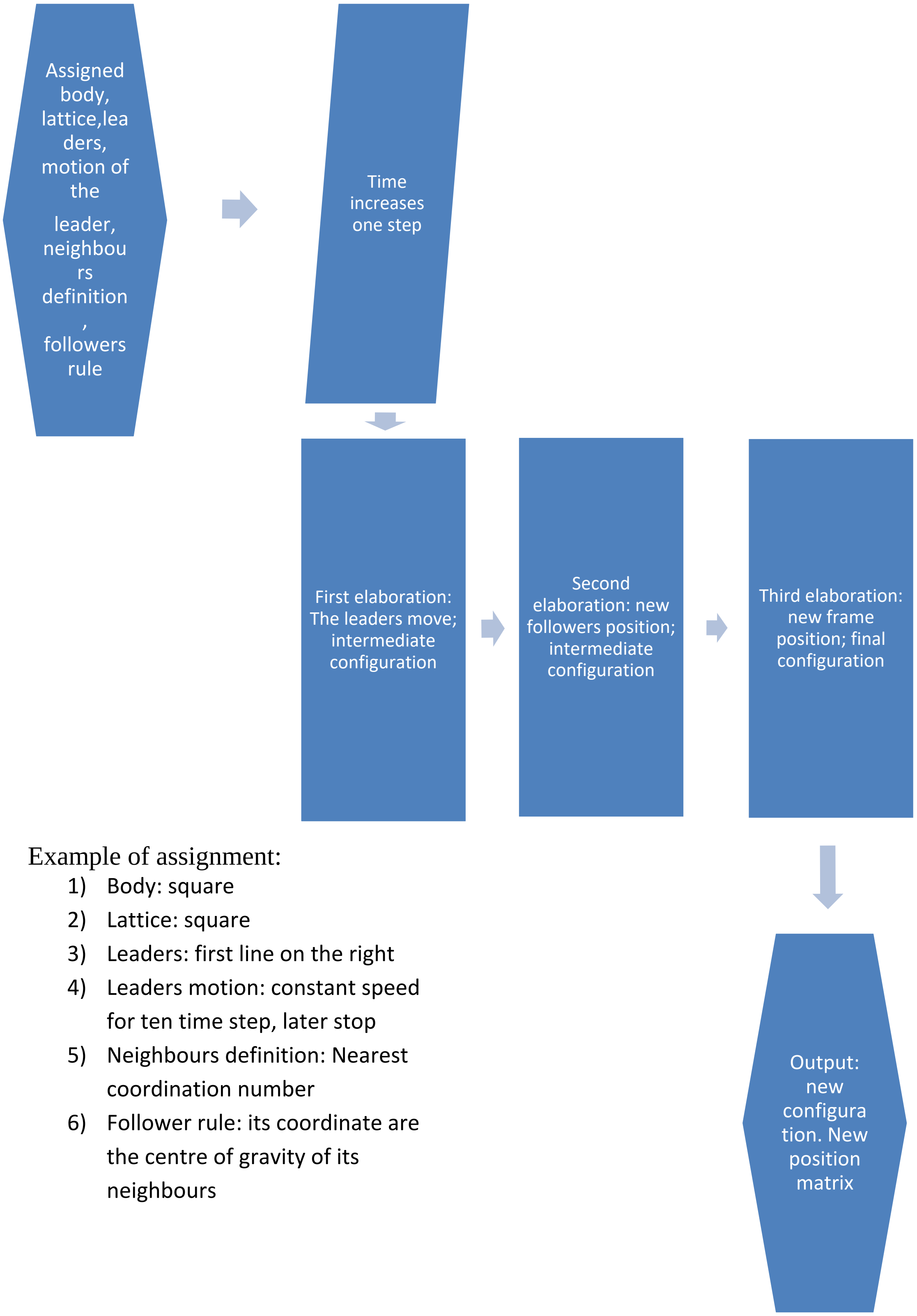

**Figure 4** Flow chart of the process

## 5. The algorithm in fracture case

To manage fracture phenomena we assume the interactions are decreasing with increasing distance between particles. Therefore, when Euclidean distance between points is “great”, they lose their interaction. To address the problem we start simply considering a threshold effect between neighbour elements, so that when the distance overcomes the threshold these elements are not anymore taken into account. To preserve symmetry of the Lagrangian neighbours we introduce ghost-like points called fictitious elements. They have the purpose of balancing the calculations of the point’s displacements. Where are these ghost elements posed? Our choice is to put them in a position which is able to recover the original shape of the lattice. All the properties of these fictitious elements are the same of the followers but their motion is not considered, because they are not in the list of the followers. They are just in the right position to balance the cell. As we have seen [6] a change in their position produces effects such as the contraction or loosening of the lattice in the deformed configuration. In fact, varying the distances of the fictitious elements after fracture from the true elements, plastic-like and elastic-like behaviours can be obtained. As elastic behaviour in fracture we mean the property of the fracture edges or of the disconnected pieces originated after fracture has occurred, to recover its original shape. The algorithm can be easily generalized to second gradient by introducing two different thresholds for the two shells of neighbours.

## 6 Approached types of problems

In the previous paper [6] we investigated some easy applications of a first version of the algorithm. Now we have reformulated the whole procedure, rewritten the code and in this first paper we approach some bidimensional problems of a simple shape object subject to imposed strain of some leaders which are significantly interesting to show the coherence of the model and its adaptability in showing different physical phenomena by changing some parameters. In particular, in this paper, we analyze:

a) Simple strain and release; tensile test of rectangular shape specimen.
b) The importance of the rules: the Poisson effect.
c) The importance of the gradient.
d) Simple fracture case
e) Saint Venant principle of local action verified
f) The importance of the lattice on different way of fracture.
g) The importance of the rules for the frame evolution.

The features of each test can be summarized in Table 1 like this:

| **Parameter description** | **Value** |
|---|---|
| *Test description* | Tensile test with relaxation |
| *Lattice type and number of points* | Square (10x10) |
| *Step time of motion* | 10 |
| *Step time total* | 400 |
| *Neighbours concept* | Coordination number of the lattice |
| *Gradient* | First |
| *Interaction rule for followers* | Position as centre of gravity of the neighbour |
| *Use of weight distance* | No |
| *Limit distance for fracture* | No fracture |
| *Distance where the fictitious are placed* | Not applicable |
| *Leader points* | First column on the left clamped. First column on the right up to time step 10; after they became followers |
| *Leader velocity* | Constant velocity in positive x direction (0.35 unit/step) for the column on the right up to time step |
| | 10, 0 later |

**Table 1** Description of the test

For every test we shall show and discuss the movement of the particles, the XY movement of a significant particle (if present) and some pseudo energetic considerations by PE1 or PE2.

## 7 Numerical results

Case a) Simple strain and release; tensile test of rectangular shape specimen
The first numerical simulations will concern the behaviour of the system described in Table 1. We are considering a square sample undergoing strain from one side (the other side is clamped) at constant velocity in x direction (speed 1,2 unit/step time). At a certain time the pull is released and the leaders return to original configuration (they have changed category and are followers) attracted by the other points. So far the leaders start moving with constant velocity up to a certain time; later they became followers subjected to the rule like the other followers. The simple rule, governing followers motion is that every point must be placed in the barycenter of its neighbours; the neighbours are determined by the coordination number of the lattice; therefore the leaders motion implies a displacement of the first layer that propagates in successive time steps. This means the displacements, at each time step, involve a larger shell of points until to regards all the lattice points. The neighbours points are determined by the coordination number of the lattice or, in second gradient, considering also the neighbours of the first neighbours. In Figure 5 we can see the configuration of the lattice over different time together with the PE1 contour plot.

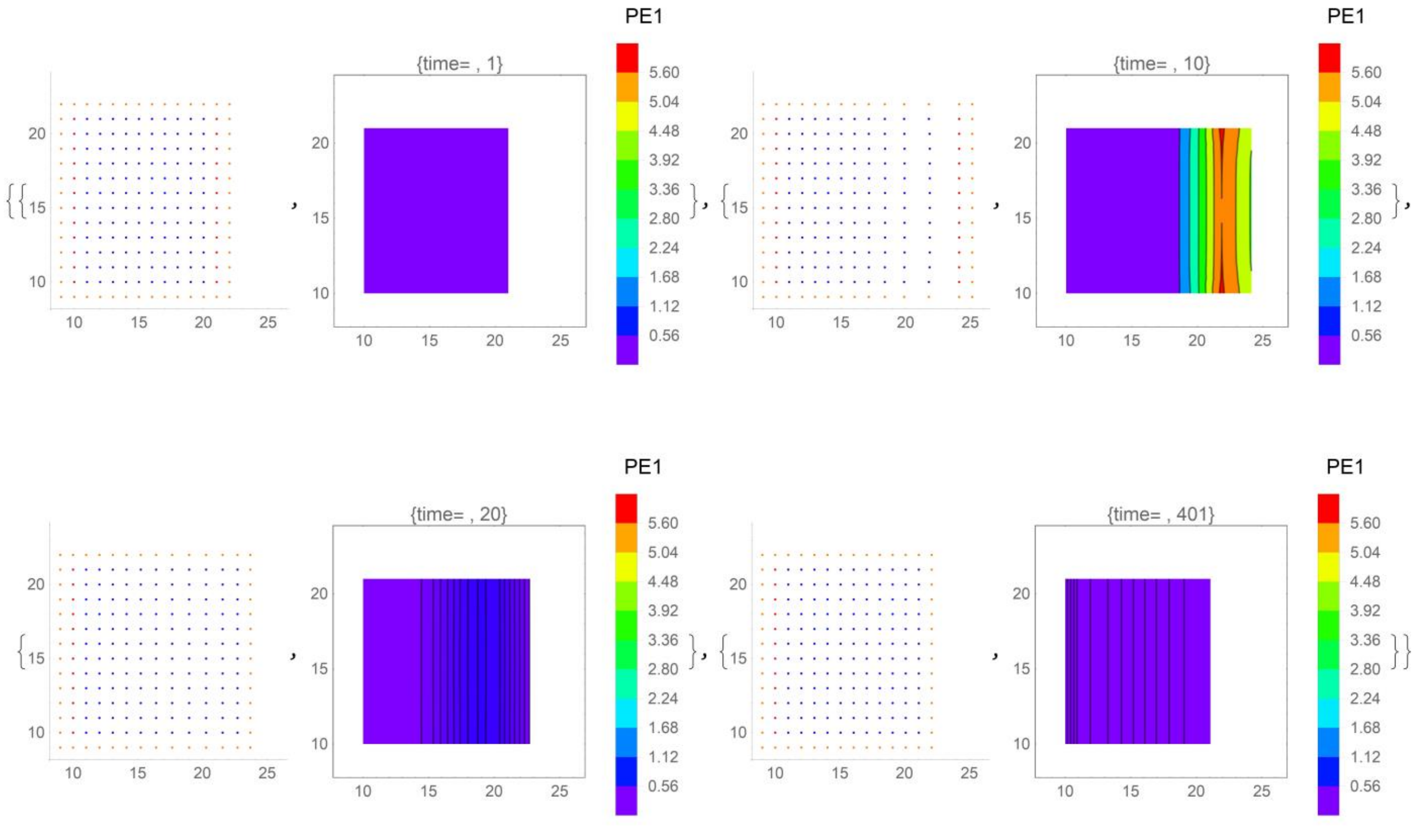

**Figure 5** Configuration of the lattice over different time (1,10, 20 and 401) and PE1 contour plot

From the figure we can outline that the x displacement of the points seems do not depend on the y coordinate; however looking at the PE1 picture we can note a light convexity that does mean this is not true. A deeper examination of the points displacements confirms as close to the frame the displacements, along x coordinate, are lower with respect to central points. This can be explained as an edge effect. In fact if we consider points on the same vertical lines those that are close to the frame follow the neighbours with a little delay owing to the different rule determining the displacement of the frame and of the followers. So they see a different situation with respect to, as an example, a central point. This effect will be more evident in other computation later. Moreover we can note as the maximum value of PE1 (red area) is not on the leader line but just on its left; this because, in this case, the leaders have in their neighbours, some points of the frame that are always close to them. This is also evident in Figure 6 where the PE2 is shown and contigue configurations are compared. We can avoid this convexity effect using a different frame or mirroring the followers to obtain an infinite sample. Finally it should be noted as at *t*=401 the lattice is not back completely to the reference configuration owing to asymptotic process of relaxation.

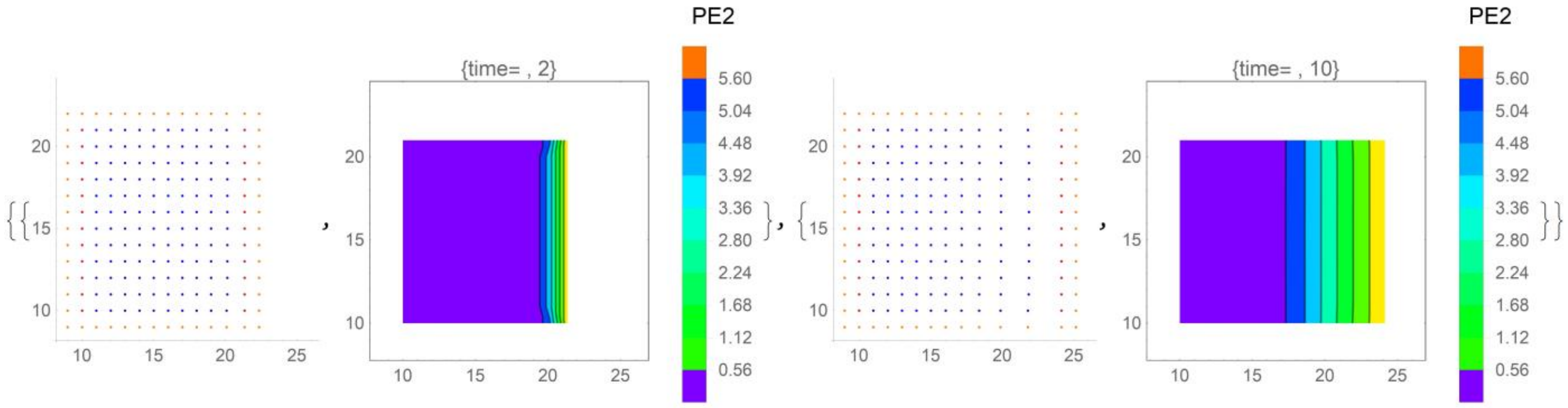

**Figure 6** Configuration of the lattice over different time (equivalent to 2 and 10 of PE1) and PE2 contour plot

Now we would like to see the evolution on time of a central point of the lattice. In Figure 8 the numbered lattice, with leaders, followers and frame are shown and in Figure 8. the behaviour of the PE1 versus time is shown for the central point j=67. Points are numbered from left to right and from bottom to up.

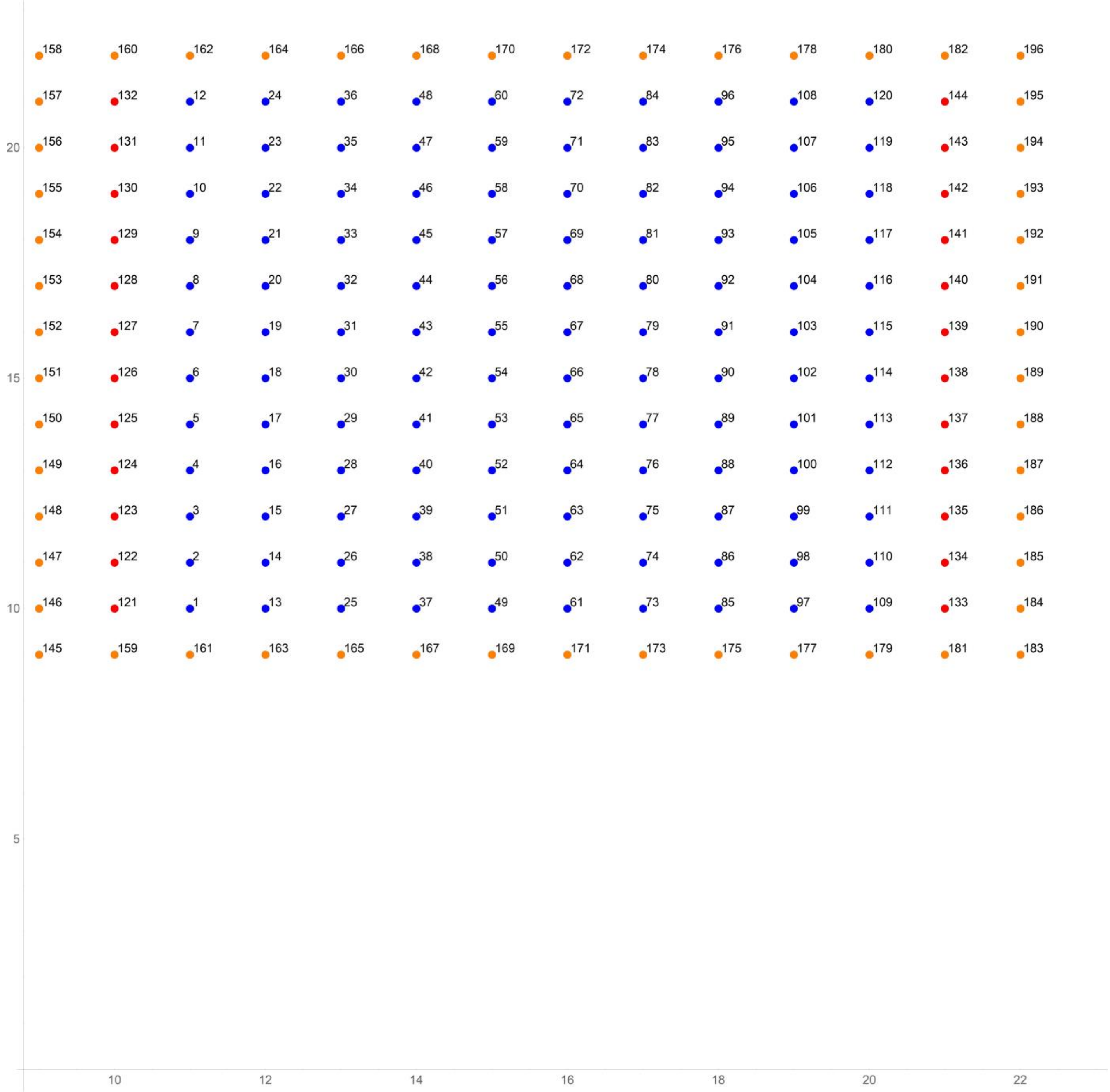

**Figure 7** Numbered lattice. Red points are the leaders, yellow the frame and blue the followers.

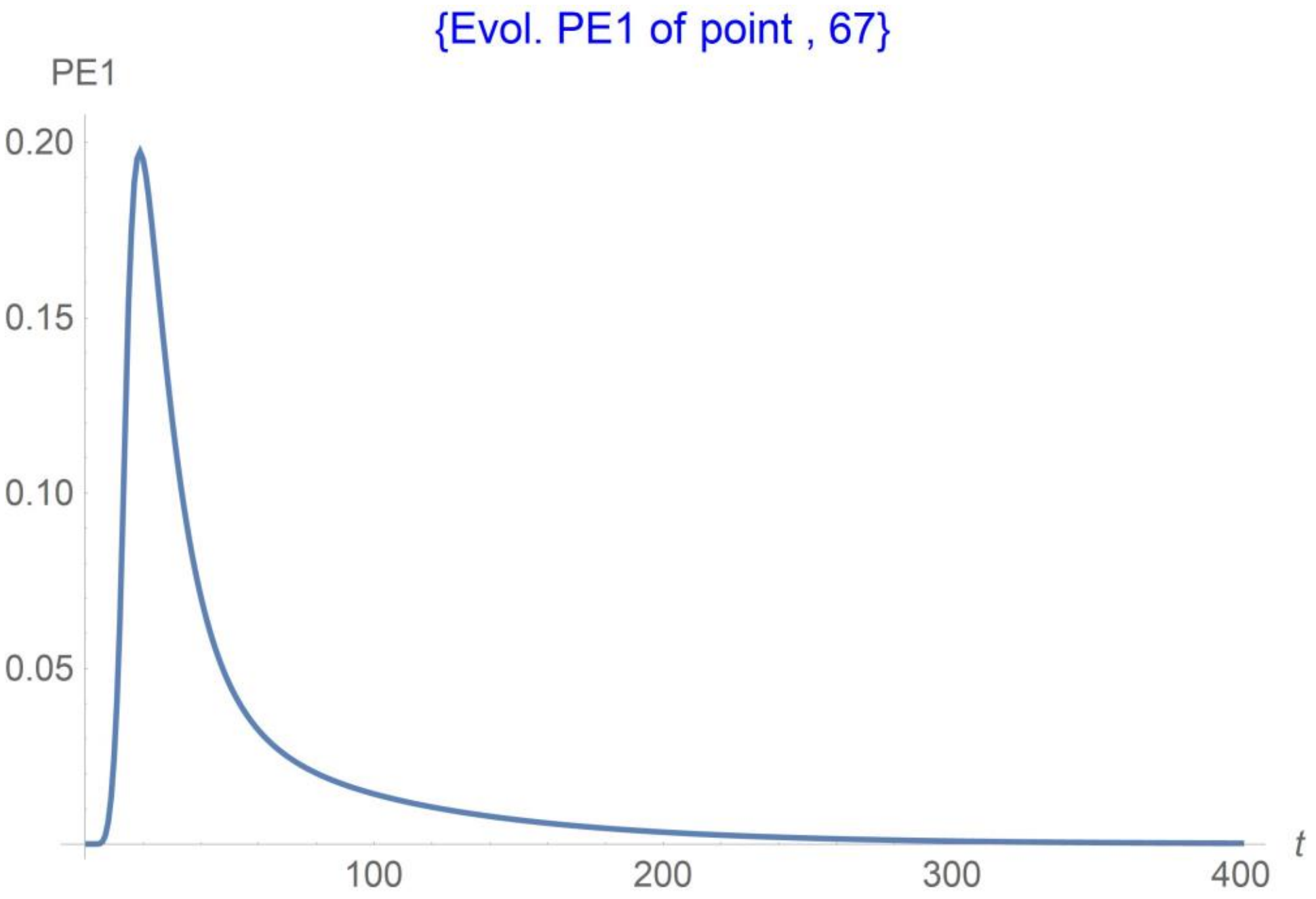

**Figure 8** PE1 versus time

The value of the PE1 increases notably when points are pulled but decrease in quite a parabolic-like way when they become followers subjected only to the rules leading to equilibrium barycentre position. This has similarity with Hooke's law where the energy is proportional to the square of the displacement. If we change point the shape of the curve remain the same but cha be less o more flared.

In Figure 9 the evolution with time of the coordinates of central point j=67 is shown. Also in this picture we can recognize the coordinate x increases linearly (velocity is constant), after a delay, owing to the propagation time, and later decrease with parabolic like behaviour to the original position. No oscillation can be observed with this kind of rule.

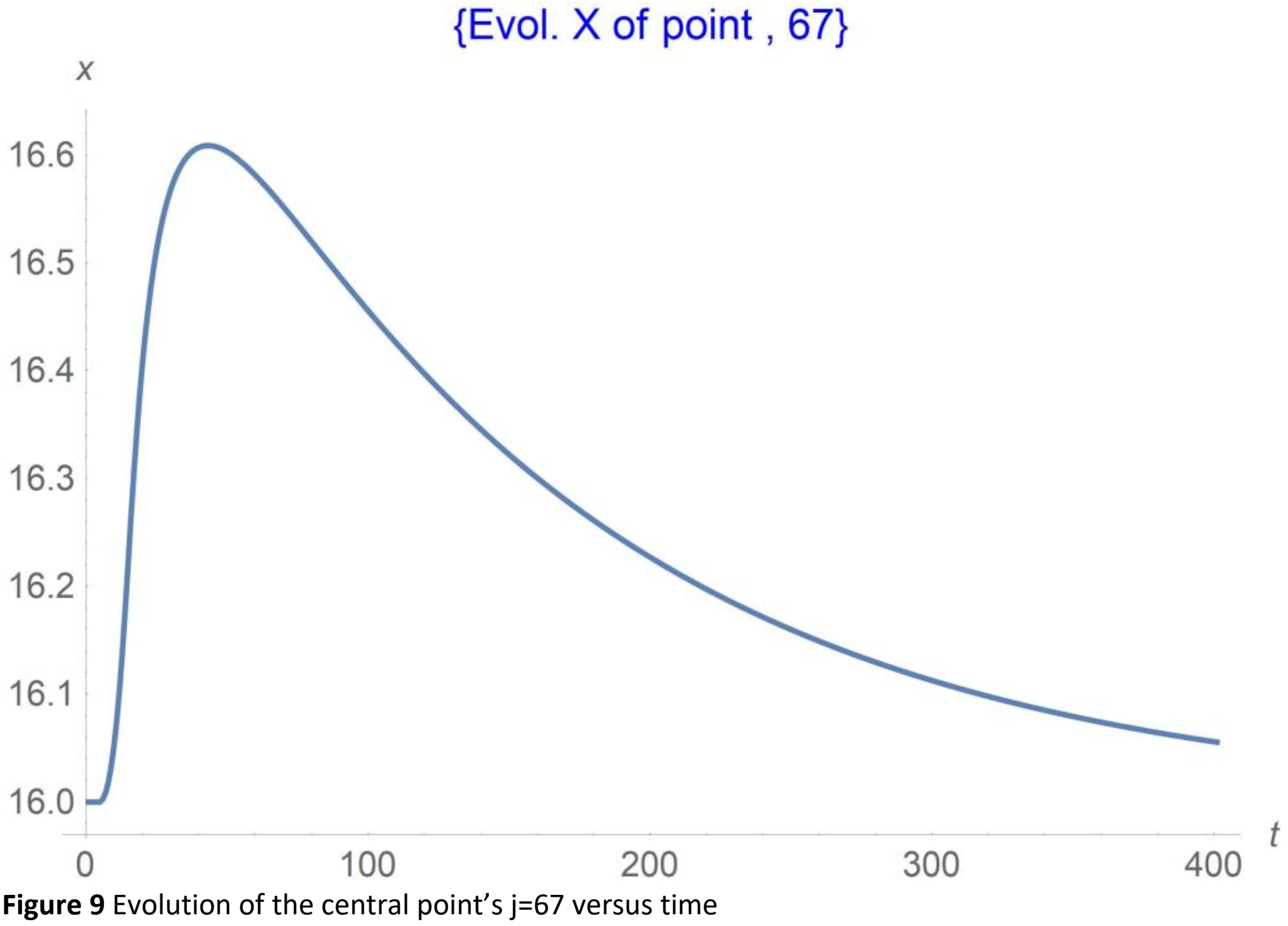

**Figure 9** Evolution of the central point's j=67 versus time

Case b) The importance of the rules: the Poisson effect

In the second case we change the interaction rules; the case is the same as a) (without release, the leaders stop after movement and the motion steps are 100) but the follower position is determined not only by a baricentric equation but the rule take into account the other coordinate. We call this method “mixed coordinate”.

In Figure 10 we can see the configuration of the lattice over different times together with the PE1 contour plot.

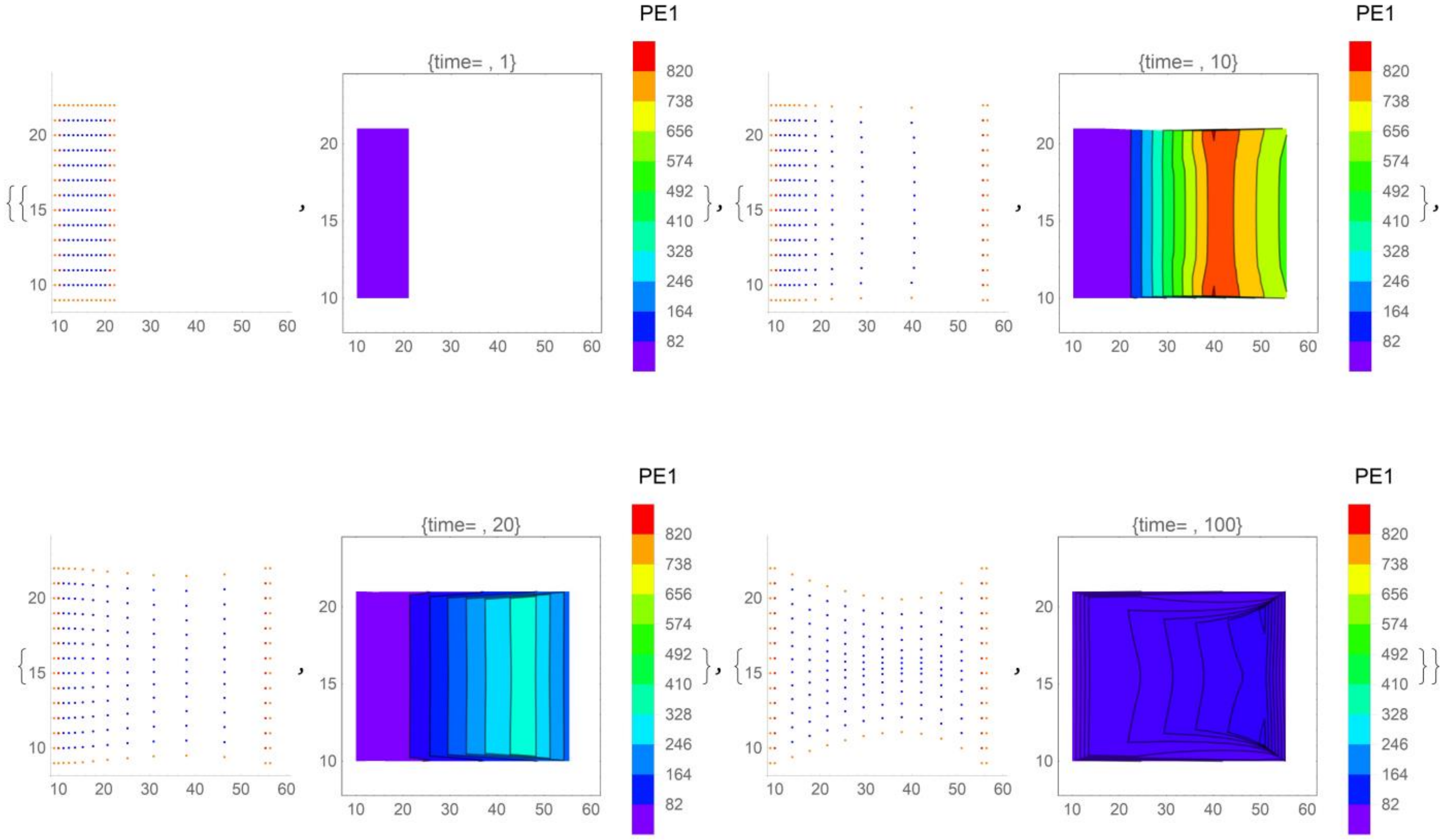

**Figure 10** Configuration of the lattice over different time (1,10, 20 and 100)  and PE1 contour plot

Lateral contraction can be seen. Note that there is a relaxation time, because the followers need time to adapt themselves. This is due to the rules expression and can be tuned as you desired. The PE1 value seems to follow the configuration; here is more evident the edge effect leading to concavity effect toward left side.

In Figure 11 the evolution with time of central point j=67, is shown and the lateral contraction, in y coordinate, is evident. The point is above the central line so their y coordinate decrease.

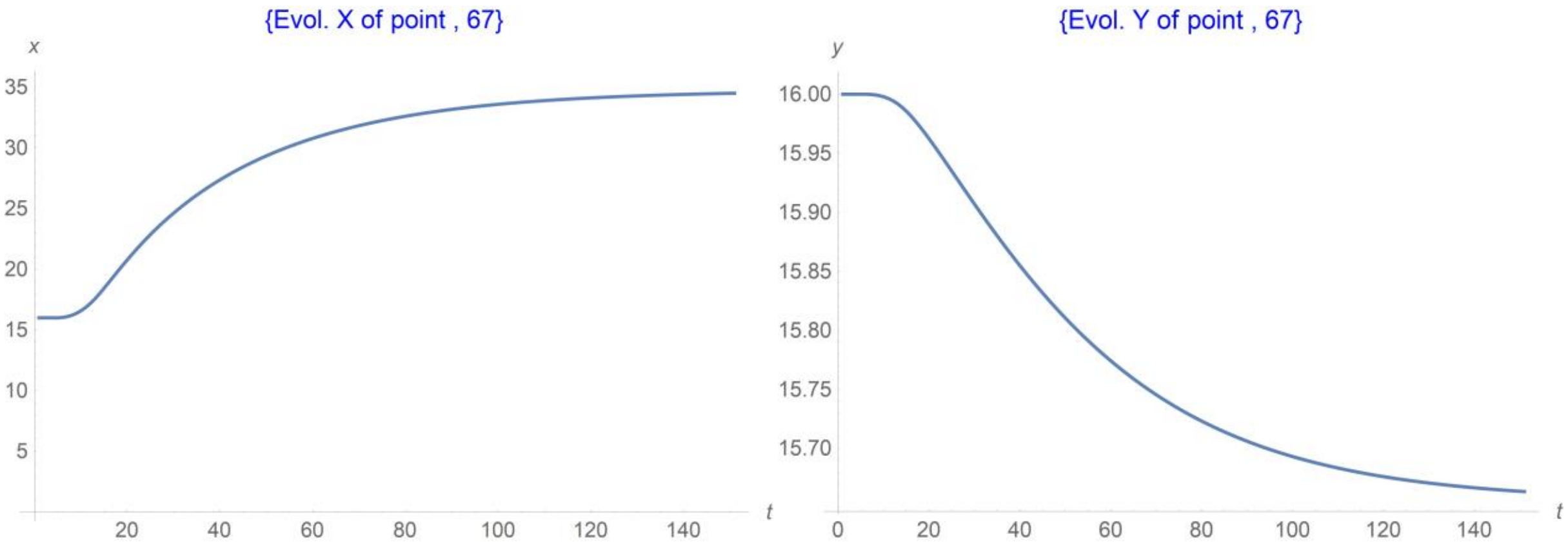

**Figure 11** Evolution of the central point j=67 versus time

Case c) The importance of gradient
The case is the same as b) but the follower position is determined by a second gradient neighbours.
In Figure 12 we can see the configuration of the lattice over different times together with the PE1 contour plot.

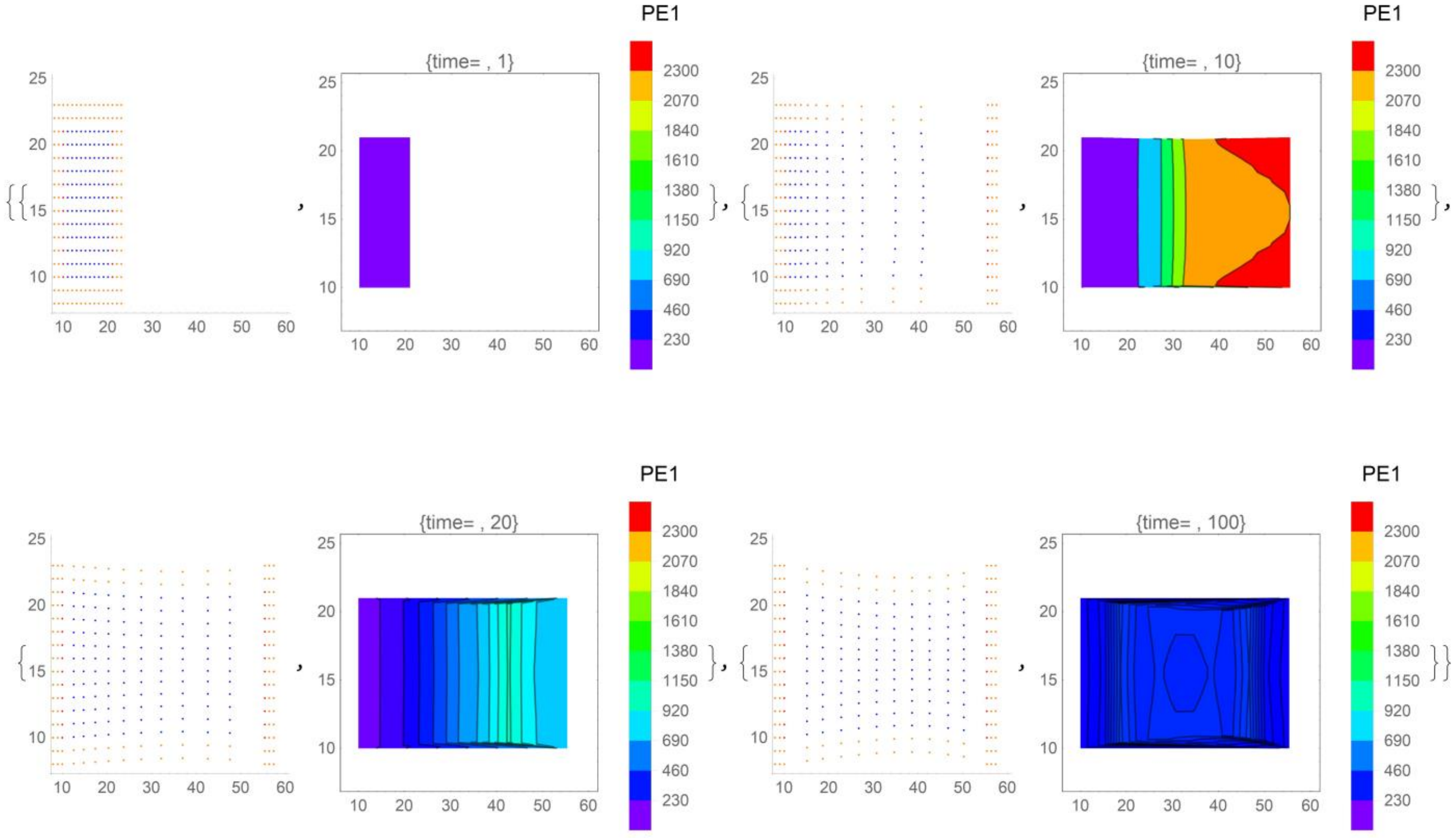

**Figure 12** Configuration of the lattice over different time (1,10, 20 and 100) and PE1 contour plot. Second gradient

Lateral contraction can be seen in a less pronounced way; this is because if we take into account a larger number of neighbours the effect is amortized. The Pe1 plot enhances the convexity of the first line of follower. We shall see better as, in the fracture case, second gradient has much more influence in some cases.
In
Figure 13 the evolution with time of the central point j=67 is shown and a lower lateral contraction, with respect the first gradient case, can be outlined and its evolution with time is quite different.

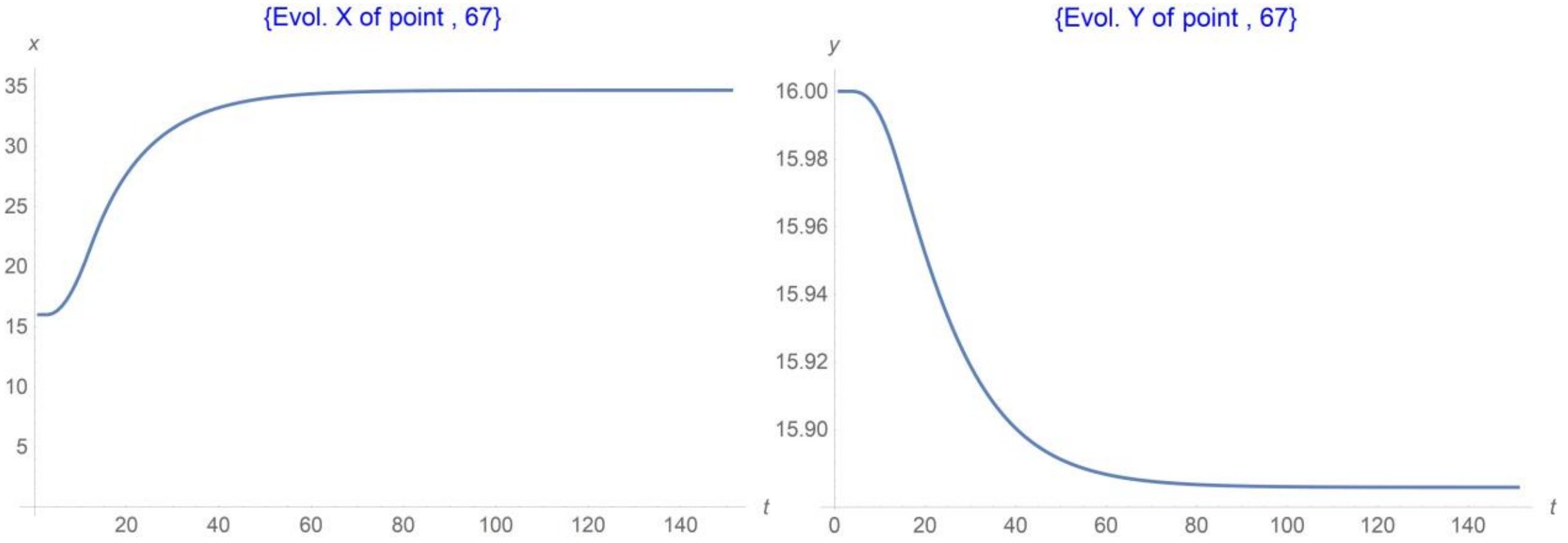

**Figure 13** Evolution of the central point j=67 versus time

Case d) Simple fracture case

A simple fracture case can investigate with the same parameter of the preceding but with the rule for follower to be in the centre of gravity of its neighbours (no Poisson case). Choosing a distance fracture of about 10 units in the following pictures we can see that the vertical fracture line is different in the case of first (see Figure 14) or second gradient (see Figure 15). Points close to the frame are detached before the others from the leaders, and this effect is more marked in second gradient case. This can be explained with the different neighbours number and also with the larger influence of the frame with respect to the first gradient case.

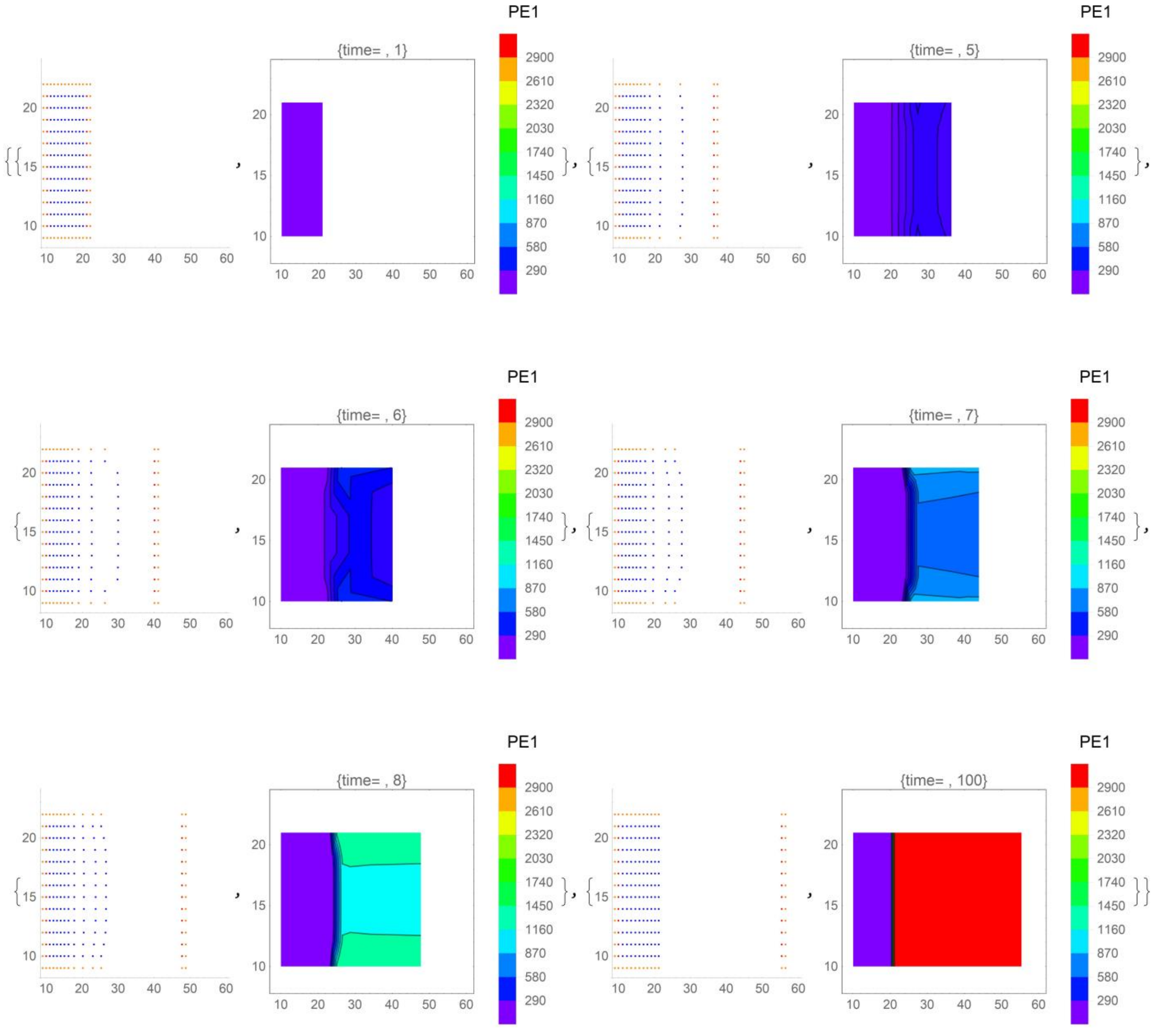

**Figure 14** Configuration of the lattice over different time (1,5, 6,7 and 100) and PE1 contour plot, fracture in tensile test. First gradient

Remember that in the fracture case the pseudo energy plot is less indicative, because we are calculating it using distance between points greater than the fracture threshold. In a future work we will consider a better definition of this parameter.

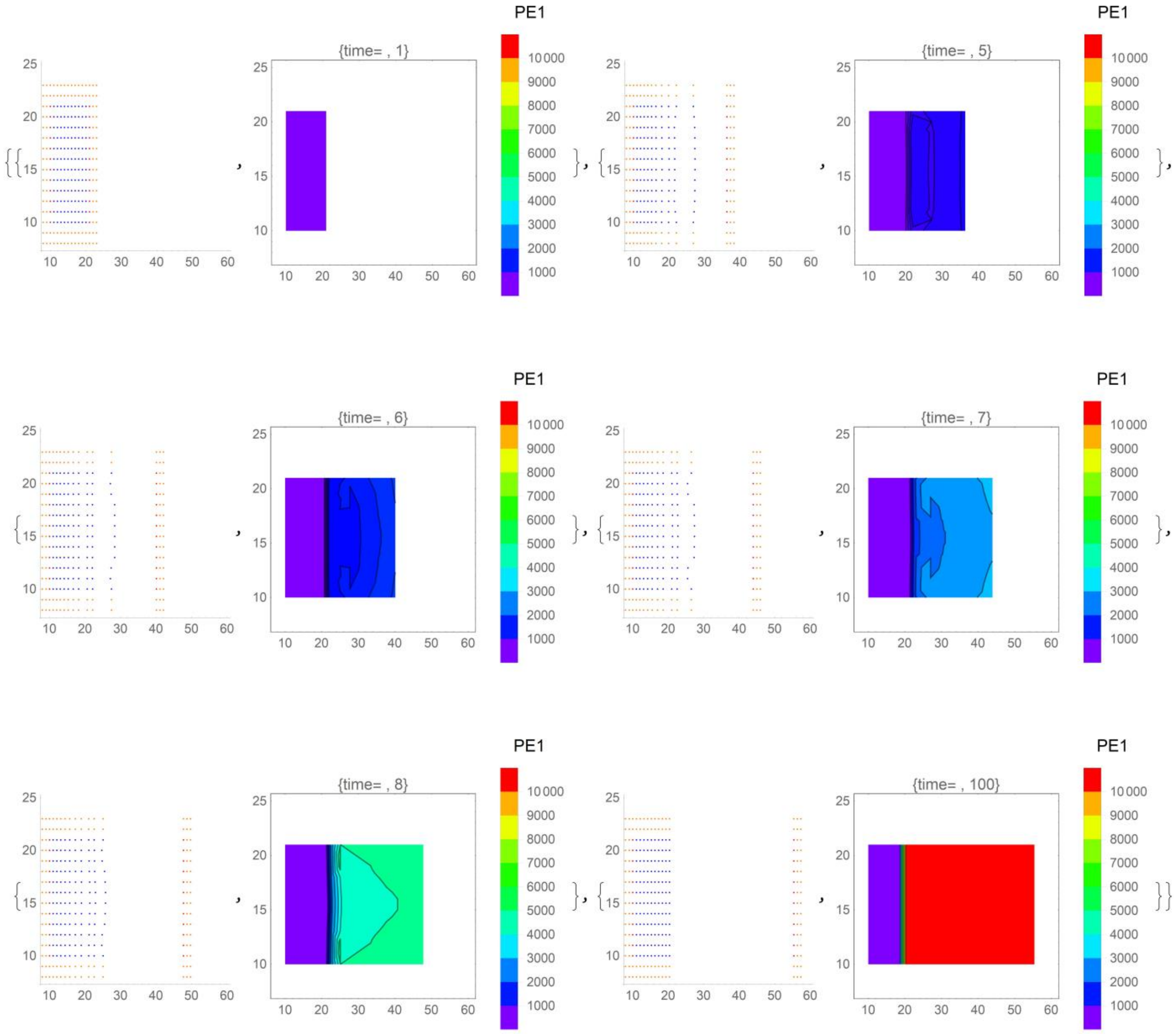

**Figure 15** Configuration of the lattice over different time (1,5, 6,7 ,8 and 100) and PE1 contour plot, fracture in tensile test. Second gradient

The importance of second gradient can be outlined in Figure 16 where x coordinate evolution versus time of x coordinate of point 103 is showed. The point is situated in middle value as Y coordinate and two lines on the left of the leaders line; differently from first gradient mode a complex behaviour can be observed because after the fracture the x coordinate has a sort of rebound. This can be explained as follow. After the fracture the point try to return its initial position ( on the left), but later some fast point on its right try to deviate it to the right. When the group is compacted they go back all together to initial configuration. So far change in the parameters can lead to complex evolution behaviour of the lattice

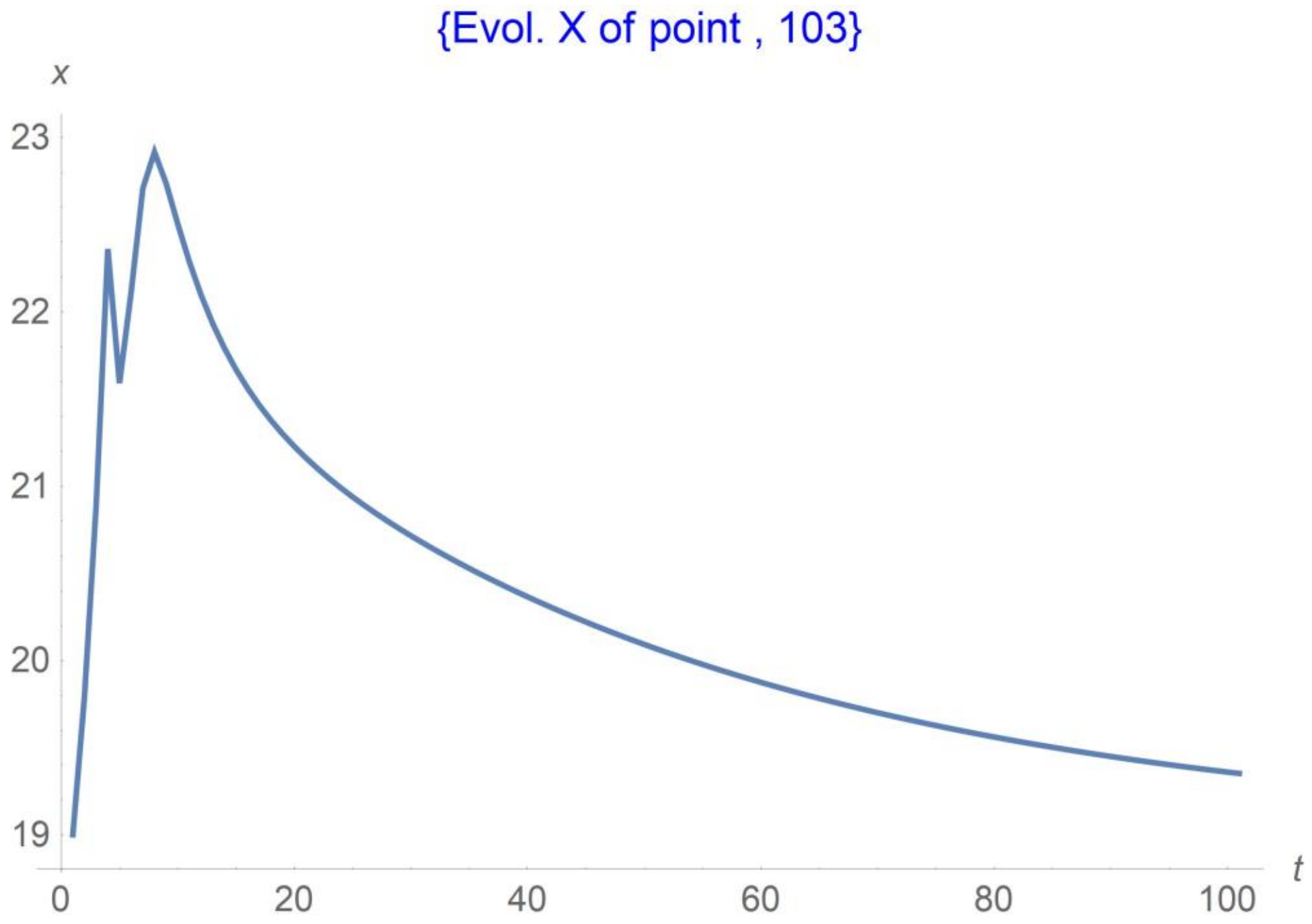

**Figure 16** Evolution of the second follower line point j=103 versus time

Case e) Saint Venant respect

In this case we shall verify the local action principle, whereby the tension at one point is not influenced by the motion of the external particles to an arbitrarily small circle of the particle in question. To this aim we consider four internal points that diverge from their initial configuration; practically we choose four internal points as leaders with opposite movement along the bisectors of the corners.

In Figure 17 we can see the configuration of the lattice in different time together with the PE1and PE2countour plot.

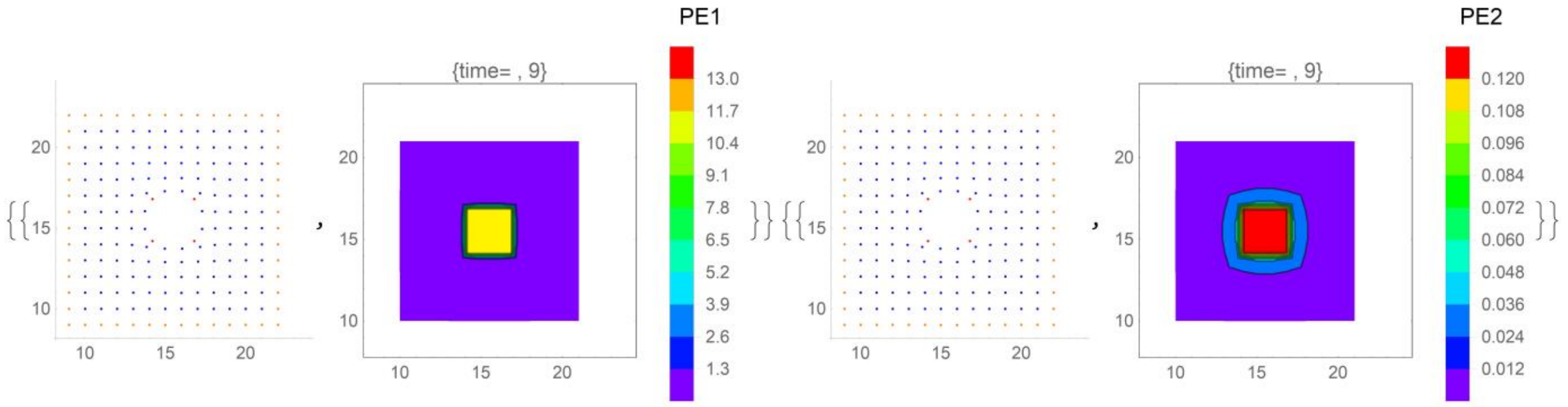

**Figure 17** Configuration of the lattice at time (9) with PE1 and PE2countour plot

As can be seen remote particles are not interested in what is going on close to the leaders.

Case f) The importance of the lattice on different ways of fracture.

The fracture mechanism is strongly dependent on lattice characteristics. In the next case we choose a hexagonal lattice, instead of square; the coordination number, i.e. the number of neighbours is also changed, being a characteristic of the lattice. In Figure 18 we are considering the fracture case for the hexagonal lattice (all parameters are the same as the square lattice fracture case). The behaviour is very different from the square lattice case as we expected. It can be noted that a point of the frame remains in the middle of the displacement. This can be explained as follows. The rules regarding the frame are simple; each point of the frame is linked to an assigned follower and its displacement from time t to t+1 is the copy of the follower. However in some cases the followers assigned to one point of the frame could be more than one. In such cases we can choose to take one of them or to consider the displacement of the point as the average value of the displacements of all its followers linked to it. This is the reason that two points of the frame remain in the middle: they are stressed from two opposite sides. Look at Case g) for a different frame rule. The behaviour is very interesting; it can be noted that a different equilibrium configuration is reached because the frame is changed and the fictious are not in the list of the followers. In the case of Figure 18 this results in a concave final surface, owing to the modified frame.

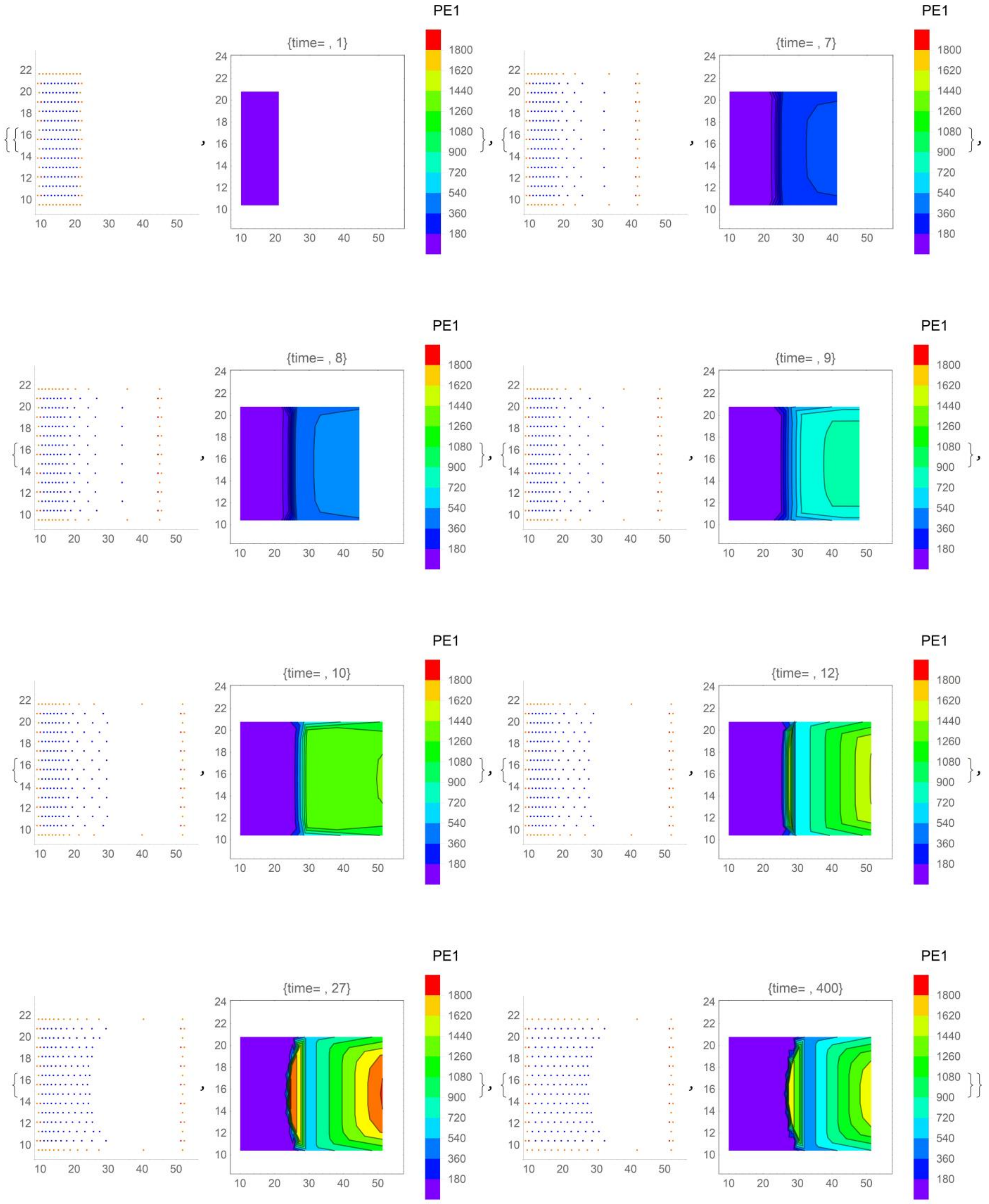

**Figure 18** Configuration of the lattice over different time (1,7,8,9,10,12,27 and 400) and PE1 contour plot

A rebound mechanism can be outlined from Figure19 This is typical of a hexagonal lattice; you can see that in the square lattice this behaviour does not exist.

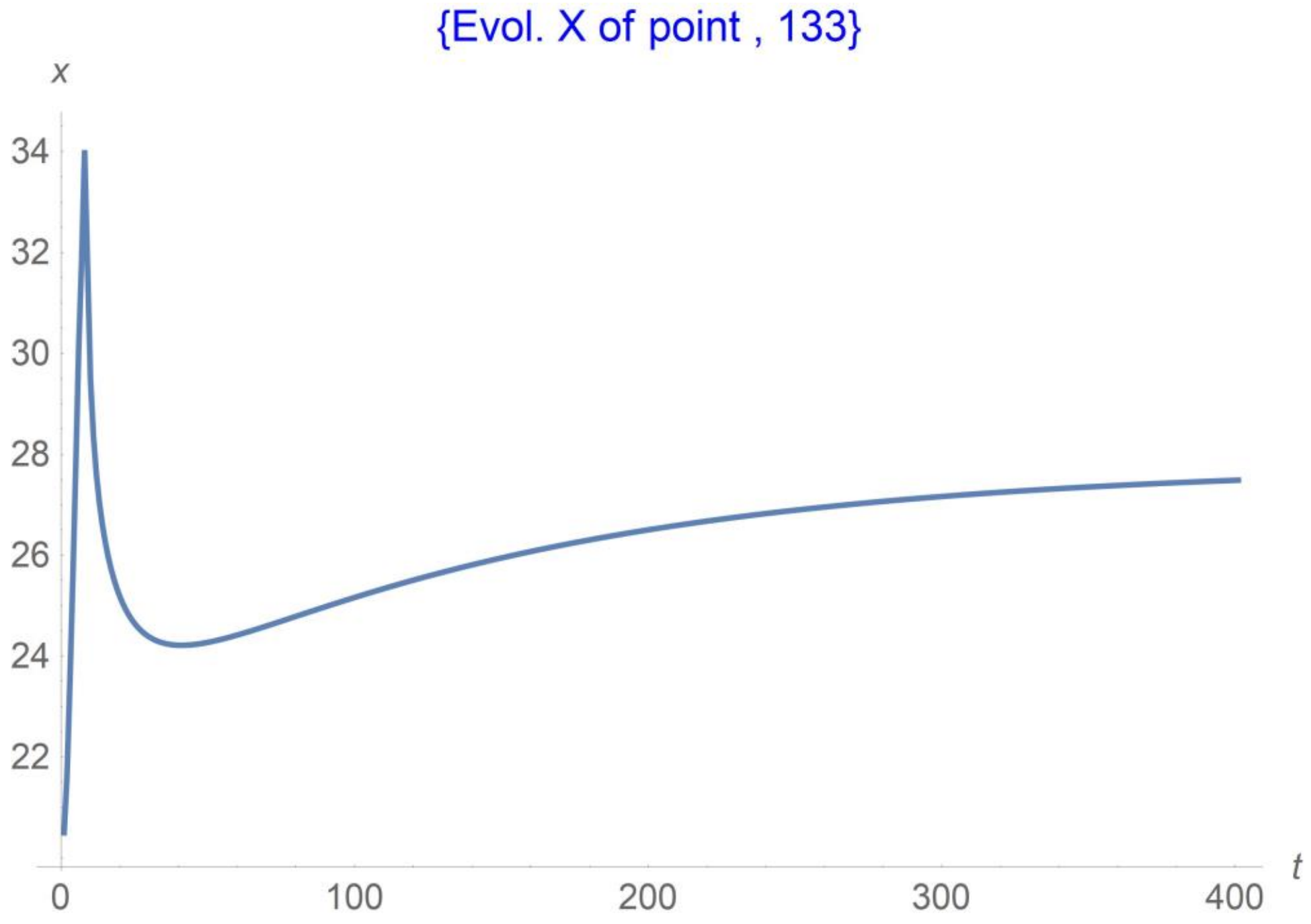

**Figure19** Evolution of the central point j=133 versus time

Also in this case second gradient mode attenuates the effects, see Figure 20. No rebound (Figure 21) can be outlined and the equilibrium configuration, after the fracture is similar to the original. Cuspidal point in X coordinate, after about 100 step, can be attributed to a second fracture happening.

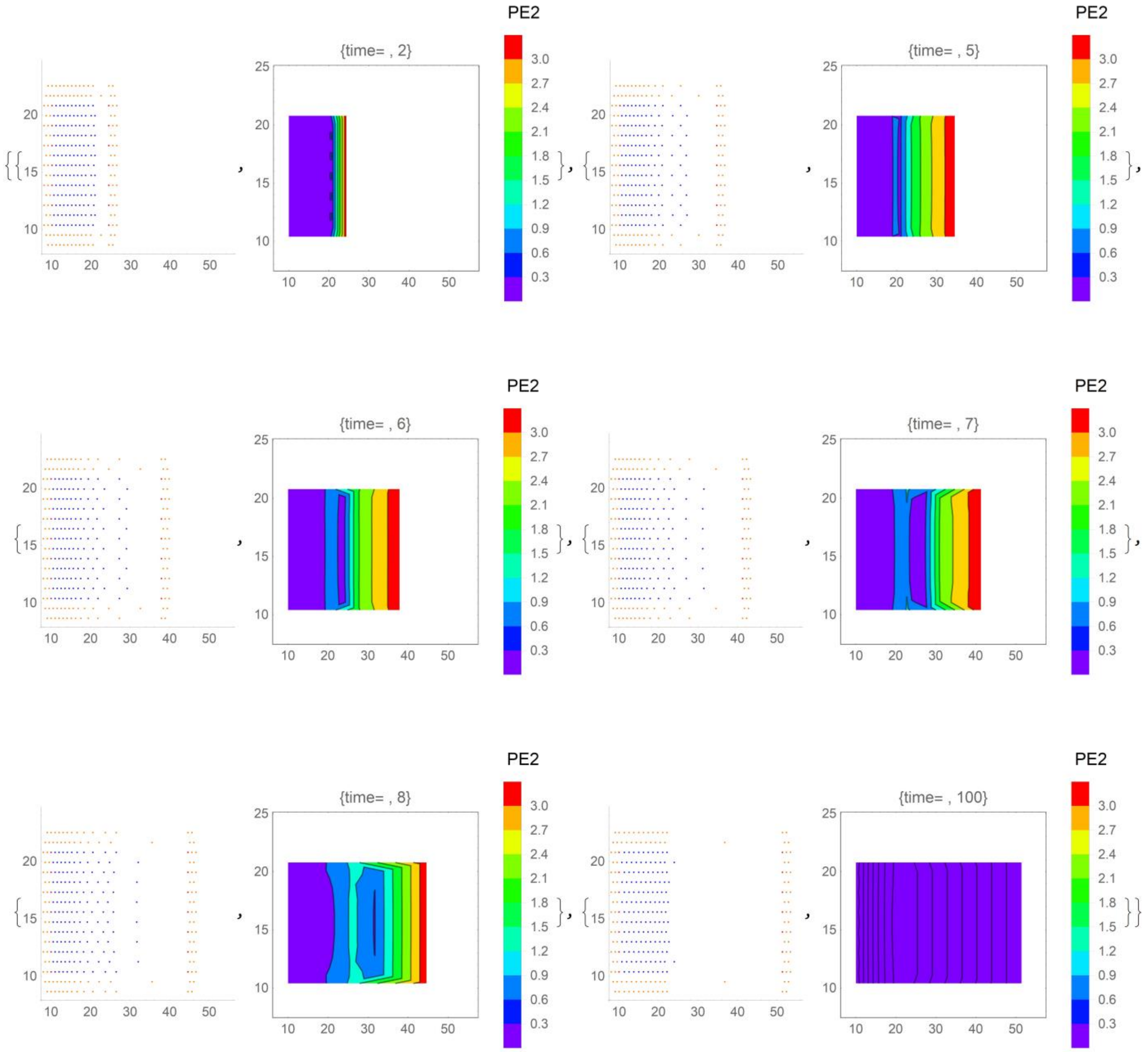

**Figure 20** Configuration of the lattice over different time (1, 7, 8, 9,10,12,27 and 401) and PE1 contour plot

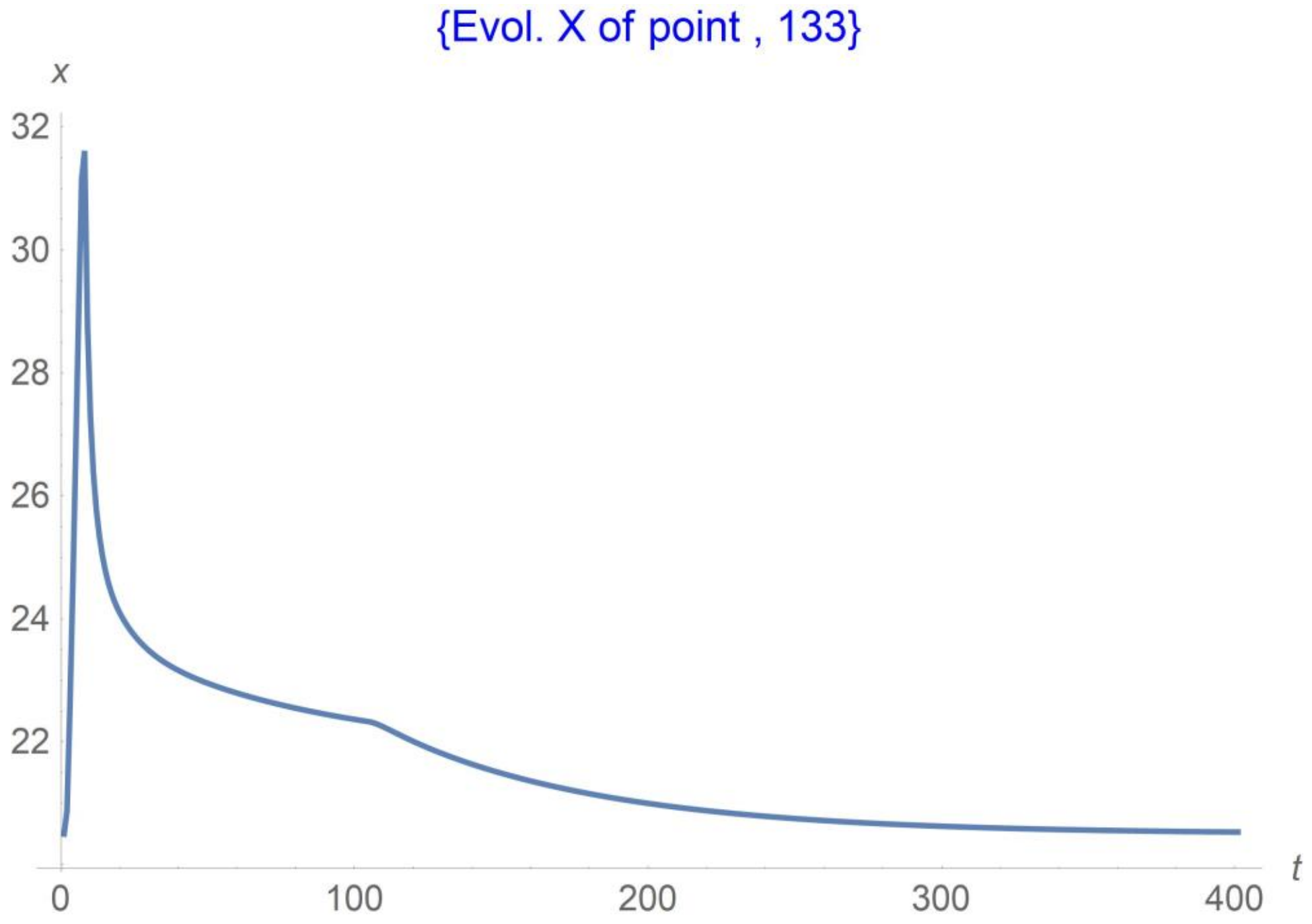

**Figure 21** Evolution of the central point j=133 versus time

Case g) The importance of the rules regarding frame motion.
We have just noted that for the hexagonal lattice there were two frame points that remain in the middle. To stress the importance of the rules also for the frame now we are considering the same case (see in Figure 22) where only the first of the followers are linked to each point of the frame. The behaviour is completely different and more similar to the square lattice.

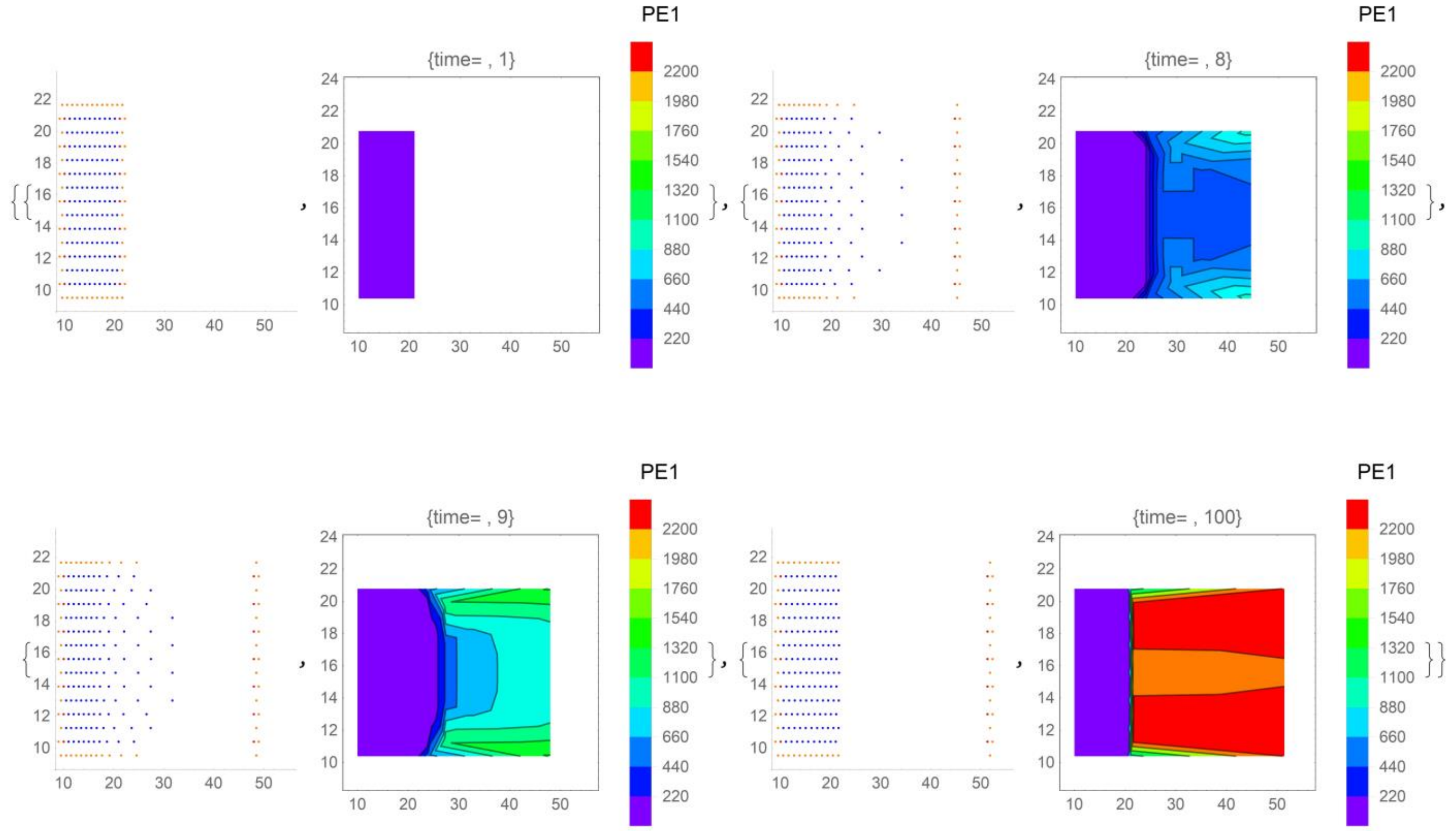

**Figure 22** Configuration of the lattice for different time (1,8, 9 and 100) and PE1 contour plot.

Also the convexity of the surface fracture is changed with respect to the preceding case (second gradient with average value for the frame) being similar to the square lattice case. This is because there are no more points of the frame in the middle and the final configuration can return close to the initial.

## 5. Conclusions

In this work we have presented some numerical simulations regarding strain deformation of a continuum medium. We started from an initial configuration imposed a strain and calculated the deformed configuration, using a position based dynamics method able to take in account complex physical effects. This means that deformation configuration is calculated not by Newton law but only by the relative positions between the particles of the system and the characteristics of the lattice. In a preceding work we verified the results of this tool are in accordance with results obtained by FEM, also in the fracture case; so far, in this paper, we have tried to generalize its application. We have showed a computational tool able to describe the behaviour of different materials, by changing some parameters of the algorithm. Computational costs are low because we do not solve differential equations but only algebraical equation systems. Working with a transformation operator between matrices the job can be parallelized between the GPU cores of the powerful video card. This is not a new kind of physics, just a graphic representation of a plausible behaviour; keep in mind that, up to now, you do not start from the constitutive equations of the materials leading to the rules governing points displacement. Actually we just imitate a known behaviour adjusting the algorithm parameters. Anyway the results are alsointeresting even if still in a preliminary form; in a next paper we shall try to connect the rules of our model with physical proprieties of the material. Pseudo energetic considerations are introduced to describe different deformation regimes, such as elastic and plastic and to achieve a better understanding of the process. This is preliminary to introducing potential descriptive interactions depending on the relative distance between the particles, which are able to reproduce the well known physical behaviour. Many questions remain open. How stable and robust is the model? What is possible to reasonably describe with this model and what are the physical reasons of its success? Is there a hidden dynamic? Does a connection exist between pseudoenergy and a real potential? Does a physical meaning exist (like Hooke's law) of the position rules? How are the constitutive equations linked to the mixed interaction position rules, used to describe the Poisson effect? What kind of rules do we need for oscillation? Finally, the mathematical study of the homogenization of lattice systems like the one here considered seems to pose interesting problems, and will probably require non-trivial ideas in the field of functional convergence [60-65]. These, and many others, are the object of a next paper, together with a generalization in 3D. In order to meet the richness of behaviour of different materials, including potentially complex biological tissues [66-71], some other interesting behaviours are under study.